\documentclass[%
twocolumn,
superscriptaddress,
nofootinbib,
 amsmath,amssymb,
 aps, prd
]{revtex4-2}

\usepackage{xcolor} % Remove after notes are no longer needed
\usepackage[normalem]{ulem} % Remove after notes are no longer needed
\usepackage{lipsum} % Remove after filler text is no longer needed
\usepackage{soul}
\usepackage{ulem}

\usepackage{mathtools}
\usepackage{subcaption}
\usepackage{graphicx}
\usepackage{amsmath}
\usepackage{hyperref}
\usepackage{booktabs} % for making pretty (read: more easily customizable) tables
\usepackage{makecell} % for formatting cells (to be multiline) in tables
\usepackage[toc,page]{appendix}

\DeclareSymbolFont{cmletters}{OML}{cmm}{m}{it}
\DeclareMathSymbol{v}{\mathalpha}{cmletters}{"76}

\begin{document}

\title{The impact of binary neutron star outflow mass models on kilonova parameter estimation}

	\author{Amelia Henkel}
	\affiliation{Department of Physics and Astronomy, University of New Hampshire, 9 Library Way, Durham, NH 03824, USA}

   	 %\author[0000-0003-4617-4738]{Francois Foucart}
   	\author{Francois Foucart}
	\affiliation{Department of Physics and Astronomy, University of New Hampshire, 9 Library Way, Durham, NH 03824, USA}

	\author{Selah Melfor}
	\affiliation{ASTRON, Netherlands Institute for Radio Astronomy, Oude Hoogeveensedĳk 4, 7991 PD, Dwingeloo, The Netherlands}
    \affiliation{Anton Pannekoek Institute for Astronomy, University of Amsterdam, Science Park 904, 1098 XH Amsterdam, The Netherlands}

    \author{Samaya Nissanke}
	\affiliation{Anton Pannekoek Institute for Astronomy, University of Amsterdam, Science Park 904, 1098 XH Amsterdam, The Netherlands}
	
	\author{Alexandra Wernersson}
	\affiliation{Anton Pannekoek Institute for Astronomy, University of Amsterdam, Science Park 904, 1098 XH Amsterdam, The Netherlands}
	
	\author{Uddipta Bhardwaj}
	\affiliation{Anton Pannekoek Institute for Astronomy, University of Amsterdam, Science Park 904, 1098 XH Amsterdam, The Netherlands}

\begin{abstract}

The ejecta from binary neutron star mergers, which powers its associated kilonova, can inform us about source properties, merger dynamics, and the dense nuclear matter equation of state.  While now in the era of multi-messenger astronomy, we remain more likely to observe purely electromagnetic (or purely gravitational-wave) signals due to the duty cycle and maximum observing distance of gravitational wave detectors.  It is thus imperative to be able to perform high-accuracy parameter inference of purely electromagnetic detections.  In our previous work we studied a suite of semi-analytical formulae for the masses of the dynamical ejecta and disk ejecta from binary neutron star systems to measure how their predictions compared across a broad parameter space of possible mergers.  Here, we use that same collection of ejecta formulae in an end-to-end analysis to generate mock multi-wavelength kilonova signals and recover the intrinsic merger parameters.  By generating mock light curves for a broad range of possible mergers and with a variety of ejecta models, and then performing parameter estimation with different ejecta models, we measure how reliably and consistently we can use the ‘map’ between intrinsic and outflow properties provided by the formulae to gain source (intrinsic) information from purely electromagnetic detections (observables).  We find that the posteriors probability densities are prone to biases when we vary the choice of reference model for our sampler, especially away from the best studied region of the binary parameter space, i.e. near equal mass, low mass binaries. This highlights both the need for improved models for the mass of the ejecta, and the need to exert caution when performing parameter estimation with a single mass model. We also find that most models predict that parameter degeneracies in kilonova light curves are largely orthogonal to those measurement observables, such as the chirp mass, from gravitational wave signals, indicating that kilonovae may provide a lot of additional information even in multi-messenger observations, if better modeled. 
%Our sampling algorithm optimizes fidelity to the injected ejecta mass and velocity, and when we introduce a different relationship between intrinsic and outflow parameters, we obtain best-fit intrinsic parameters which have been biased to map to this new relation.  
  %When the same ejecta formula is used for LC generation and for the sampler’s reference, which we call ‘direct sampling,’ the injection values are unilaterally included in the searched space  \cite{Levan_2023}
 %{\FF I'd remove the last sentence}
  
\end{abstract}

\maketitle

%\author[0000-0001-8688-5273]{Amelia Henkel}

\section{Introduction}\label{sec:introduction}

As gravitational wave detectors continue to improve by the day, and with multiple next-generation facilities on the horizon, the world of astronomy has waited with anticipation for another watershed event like GW170817/AT2017gfo.  
%While multiple candidate binary neutron star mergers have been discovered to date, it is practically much more realistic to expect a ‘single-messenger’ signal for BNSs, i.e. a GW event or a GRB or a kilonova candidate.  
In the mean time, one other binary neutron star (BNS) merger has been observed solely through gravitational waves (GW). Multiple kilonova candidates have also been observed in association with gamma-ray bursts (GRBs) \cite{Fong_2015,Troja_2018,Rastinejad_2022,Levan_2023,Hamidani_2024,Stratta_2025}, encompassing those classified as both long and short GRBs \cite{Gottlieb_2023}, without associated GW signals. 
Overall, single-messenger events appear significantly more common than multi-messenger observations.

In general, there is no guarantee that an observed GW event will be followed by an electromagnetic (EM) transient, even if we precisely localize the source region and perform EM follow-up; the system could be off-axis, or there could be no EM counterpart, for example.  On one hand, simulations have shown (see, e.g., \cite{Radice_2018}) that the presence and nature of BNS outflows are highly dependent on the neutron star equation of state (EOS), total system mass, and mass ratio; for example, simulations tend to agree that a high-mass, equal-mass merger is unlikely to produce any significant EM radiation.  On the other hand, the observing run schedule for the LIGO-Virgo-KAGRA gravitational wave detection network (`LVK') and their regular instrumental upgrade periods are such that an event could occur while GW detectors happen to be offline; indeed, this was the case for GRB230307A and its Ultraviolet/Optical/Infrared (UVOIR) transient \cite{Levan_2023,Gillanders_2023}. The distance to which we can detect GRBs, kilonovae, and GW signals are also quite different, and GRBs are only detectable for the small fraction of binaries observed slightly off-axis. In the absence of a coincident GW event, there is thus a strong motivation to develop capabilities to compare an observed kilonova to fiducial multi-wavelength models and use such observations to estimate the system’s ejecta properties. Multiple groups have independently performed such inferences for GW170817~\cite{Kedia_2023,Ristic_2025}.    
%Additionally, GRBs, which quickly localize the source region, are detectable far beyond the current GW sensitivity volume.   % {\FF Not necessarily useful for KN though, as we can't detect them very far} {\AH I'd politely argue that they are useful for KNe, as evidenced by GRB230307A!  But I'll defer to you: would you prefer I remove this text, or were you just commenting?}

There is, thus, timely motivation for constraining how observable EM signals relate to intrinsic BNS properties. Kilonova modeling remains however a difficult process: one requires detailed information about the mass, composition, and geometry of the outflows as well as an accurate modeling of the energy deposited by nuclear reactions as these outflows expand, of the thermalization of that energy within the outflows, and of the opacities of the heavy nuclei produced by these reactions. In this work, we focus solely on the question of modeling the mass of the outflows, and the influence of modeling uncertainties on parameter inference; this is, however, clearly just one potential source of error in the modeling of kilonovae. To date, a number of semi-analytical formulae exist to estimate the amount of dynamical and/or post-merger disk ejecta from BNSs as a function of source parameters such as the mass ratio $q$ or reduced tidal deformability $\tilde{\Lambda}$; see, e.g.~\cite{Kruger_2020,Dietrich_2017,Nedora_2021}).  These formulae are generally derived as fits to ensembles of BNS numerical simulations.  In our previous study \cite{Henkel_2023}, we examined a collection of dynamical and disk mass models and measured how their output compared across the horizon of possible BNSs.  We now wish to assess the relative behavior of these models in an applied setting, namely, by using their output as input for a fiducial kilonova model and subsequently measuring our ability to recover the true injection parameters.  By using different ejecta models for light curve generation versus as reference for parameter inference, we hope to identify any biases in the parameter recovery that would suggest the ejecta models relate 
the same observables to different intrinsic parameters.
%the same intrinsic properties to different outcomes.  

%{\FF We want a clear discussion of other sources of error in kilonova modeling, beyond what we are doing in this paper (radiation transport, detailed outflow properties, nuclear reactions and opacities). Clearly explain that we are only considering the map between ejecta mass/velocity and binary parameters with a very crude lightcurve model, to address only the question of the potential biases introduced by the use of one fit to numerical simulations over another.}

The scope of this paper is to elucidate how accurately and precisely we can recover source properties from an observation \textit{with existing relations for outflow properties}, which themselves map between intrinsic binary parameters and outflow parameters.  With a simple ansatz light curve model, we consider a suite of different outflow formulae as input, sampling our light curves while varying the outflow formula provided as reference to the sampler, in order to measure potential biases in our parameter recovery. 
%As we keep the model connecting outflow masses to lightcurve fixed, all other potential sources of error are here ignored.
%We leave an in-depth consideration of the degeneracies between outflow properties and observable properties for future work. 

This work is guided by two central science questions: 
\begin{enumerate}
\item Can we use semi-analytical BNS ejecta formulae for parameter inference to accurately identify the source properties of candidate events? Does the accuracy of the PE change measurably for different source properties?
\item What are the systematic uncertainties that are introduced, and of what magnitude, when we use different models to recover the properties of the same event?
%How do these errors compare to other sources of error in kilonova modelling and BNS PE?
\end{enumerate}

To isolate the impact of model choice on parameter estimation, we neglect other sources of error in this work.  In practice, several aspects of BNS ejecta production are uncertain, which obfuscates the map between intrinsic, outflow, and observable properties.  Uncertainties in the dense matter equation of state (EOS) impact how massive the remnant object can become before collapsing into a black hole --- which has implications for the amount and nature of ejecta produced. Neutrino-matter interactions impact the composition of the outflows (i.e., whether they are neutron-rich, or what $r$-process elements are produced), and different implementations of neutrino transport result in different predictions for the properties of the ejecta \cite{Nedora_2021}. Similarly, the inability of most modern simulations to accurately resolve the evolution of magnetic fields during merger creates biases in outflow predictions.

There are many possible sources of matter outflows in BNS mergers: shocked or tidally stripped dynamical ejecta or post-merger winds, which itself can be further sub-classified according the exact process powering the wind; and these outflows have different composition, geometry, and dependence in the intrinsic parameters of the binaries. Disentangling the contributions of different outflows can be difficult, especially without knowledge of the inclination angle of the binary.  For example, the axisymmetric models from \cite{Korobkin_2021} vary in bolometric luminosity by two orders of magnitude depending on viewing angle, and have shown that viewing angle can impact the dominant color of the kilonova \cite{Darbha_2020,Zhu_2022}. 
The diversity of potential outflows may also make it difficult to extract equation of state information even if our models are fully accurate, as shown e.g. in~\cite{Ricigliano:2024lwf}.

Finally, simplifying assumptions about the outflow composition are often made, such as assuming an average grey opacity rather than performing full radiative transport simulations with detailed line opacities. Kilonova light curves are also impacted by uncertainties about nuclear reactions and nuclear heating, as well as the properties of neutron-rich nuclei \cite{Wollaeger_2018, Radice_2018, Kedia_2023}.
%{\FF This section needs citations}
%(What is the impact of this in terms of observables?)

%We are asking these questions because we are interested in exploring the map between intrinsic and outflow parameters (an extension of our previous work \cite{Henkel_2023}).  Additionally, we wish to ascertain the accuracy and self-similarity between different semi-analytical ejecta formulae when employed in a kilonova PE pipeline, as is commonly done in the literature (cite)

Disentangling these many sources of uncertainties is beyond the scope of the current work. Here, we extend our previous work~\cite{Henkel_2023} and attempt to ascertain the reliability of semi-analytical ejecta formulae when employed in a kilonova parameter inference, in order to estimate the impact of that single source of errors on kilonova modeling. 
%{\SN do you also mean single source of error as it seems the model uncertainty encompasses many errors}

%{\SN do you envision work to impact also EM identification?} -- No, we do not!
We structure the paper in the following way.  In Section \ref{sec:methods}, we present our methodology for developing and subsequently sampling our mock light curves.  We then present our findings in Section \ref{sec:results}, with a thorough description of their interpretation.  In Section \ref{sec:conclusion}, we summarize the key takeaways of our project and elaborate upon suggested topics of future investigation. 
%{\FF Maybe point to the fact that mass models are already in use in GRB/KNe PE work, and this is thus an important source of error to assess}

%%%%%%%%%%%%%%%%%%%%%%%%%%%%%%%%%%%%%%
%%%%%%%%%%%%%%%% METHODS %%%%%%%%%%%%%%%%
%%%%%%%%%%%%%%%%%%%%%%%%%%%%%%%%%%%%%%
\section{Methods}\label{sec:methods}

\subsection{Light Curve Generation}

To produce our light curves, we begin with a grid of values for $M_1$, $M_2$, and $R$ under the assumption that the NSs have equal radii. We note that for most equations of state, there are typically only small variations in radii between most neutron stars; with the exception of objects close to the maximum neutron star mass.  With some $(M_1, M_2, R)$, we can then calculate the stars' compactnesses $C_{1,2}$, their dimensionless tidal deformability $\Lambda_{1,2}$, and the binary's reduced tidal deformability $\tilde{\Lambda}$ via the following relations:
\begin{equation*}
C=\frac{M}{R} \qquad \textrm{log}(\Lambda) = \frac{1}{2b}\left[-b-\sqrt{b^2 - 4d(a-C)}\right]
\end{equation*}
(a=0.001056, b=-0.0391, d=0.371) and
\begin{equation*}
\tilde{\Lambda} = \frac{16}{13} \frac{(M_2+12M_1)M_2^4\Lambda_2 + (M_1+12M_2)M_1^4\Lambda_1}{M^5}.
\end{equation*}
The above equations for $C$ and $\tilde{\Lambda}$ are definitions of these variables, while the relation between $\Lambda$ and $C$ is a quasi-universal relation for neutron stars \cite{Suleiman_2021}.
We obtain the velocity of the ejecta with the semi-analytical relation from Radice \textit{et al.} 2018 \cite{Radice_2018}, which is solely a function of $\tilde{\Lambda}$.  To  calculate the total ejecta mass $M_{ej}$, we assume the outflows are composed of dynamical ejecta and that 30\% of the disk mass is eventually unbound, i.e.
\begin{equation*}
M_{ej} = m_{dyn} + 0.3m_{disk}.
\end{equation*}
In practice, it is expected that depending on the merger properties, anywhere from $10\%-100\%$ of the disk mass is gravitationally unbound as ejecta, which is itself a source of uncertainty in kilonova modelling.  The choice of $\sim 30\%$ is typical of what is found in simulations of black hole-disk systems~\cite{Christie:2019lim}, but may very well be less appropriate for neutron star-disk remnants~\cite{Metzger:2014ila}.

For the masses we consider the following models: three formulae for $m_{dyn}$ --- KF, from \cite{Kruger_2020}; DU, from \cite{Dietrich_2017} ; and NEA, from \cite{Nedora_2021} --- and four for $m_{disk}$ --- KF \cite{Kruger_2020} and  NEA \cite{Nedora_2021};  DEA from \cite{Dietrich_2020}; and REA from \cite{Radice_2018}.  They are described in more detail in our previous paper \cite{Henkel_2023}, in which we qualitatively measured their variance in output to describe the same physical system.  In that work, we showed that the functional form of the model has a strong impact on the predicted ejecta mass, and that this effect was more pronounced at the `boundaries' of the BNS parameter space we studied, where fewer simulations exist and/or where the formulae had to extrapolate.  We list each of these models in \ref{table:list_of_systems}.  However, the ejecta mass (or its velocity) is not an observable property; it needs to be inferred from the kilonova.  

We employ a one-dimensional analytical model from Hotokezaka and Nakar \cite{Hotokezaka_2020}, which we denote HN, to generate our light curves.  HN does not directly distinguish between dynamical and post-merger ejecta.  Instead, they allow the user to define multiple regions of different opacity, separated in velocity space.  We chose HN for our ability to cleanly isolate the effect of different mass formulae without the additional complexity of multi-dimensional geometries.  We assume here a two-component model, with the faster ejecta being less opaque. 

Under that assumption, the HN model accepts ten parameters: $M_{ej}$, the total ejecta mass; $v_{ej}$, the average ejecta velocity; $n$, the velocity profile’s power-law index
%, which ranges from 3-5 (check)
; maximum and minimum velocities $v_{max}$ and $v_{min}$; time step $dt$ and maximum time $t$; $\kappa_{1}$ and $\kappa_{2}$, opacities which correspond to the high- and low-velocity ejecta components, respectively; $v_{\kappa}$, the demarcation region in velocity space between different opacity regions.  We hold $dt$, $t_\textrm{max}$, $n$, $\kappa_\textrm{1}$, $\kappa_\textrm{2}$, $\alpha_\textrm{max}$, $\alpha_\textrm{min}$, and $\beta_{\kappa}$ fixed, as we are interested in isolating model-dependent effects from the outflow formulae. Specifically, we hold $\kappa_{1}=0.5 cm^2/g, \kappa_{2}=5 cm^2/g, v_\kappa=0.2c, n=2, v_{max}=2v_{avg}, v_{min}=0.8v_{avg}, \textrm{t}_{max}=10 \textrm{days}, \textrm{dt}=0.01 \textrm{day}$; i.e. a hotter/less neutron rich fast ejecta and a cooler/more neutron rich slow ejecta which could reasonably be respectively identified with e.g. a shocked dynamical ejecta and a viscously-driven wind -- although the physical process behind the production of the ejecta is left unspecified here. For all the light curves we report on, we assign a constant observational error of 0.1 magnitude --- in practice, a fairly conservative estimate for observations from a distance of 40 Mpc.

This study is performed in \texttt{GEMMA}, a Gravitational waves and ElectroMagnetic counterparts Multimessenger Analysis tool originally developed by Geert Raaijmakers \cite{Raaijmakers_2021}.  Developed in Python, this software contains functionality for forward and backward modelling of multimessenger observations of BNSs and BHNSs.  \texttt{GEMMA} can be used to infer a merger's outflow properties from intrinsic parameters and vice versa, as well as predict gravitational waveforms or kilonova light curves from a set of source properties.  It is integrated with the parameter inference pipelines \texttt{bilby} \cite{bilby_paper} and \texttt{PyMultiNest} \cite{Buchner_2014}, which we employ for our analysis.

\begin{center}
\begin{table*}
\begin{tabular}{@{} *5l @{}} \toprule
 $M_1 [\textrm{M}_{\odot}]$ & $M_2 [\textrm{M}_{\odot}]$ & $R$ [km]  & Models Studied (dyn ej, disk ej for inpt)-(dyn, disk ej for recovery)    \\ \midrule
  1.8 & 1.3 & 13.0 & NK-NK, DN-KR, NN-NN, NK-KD, KK-KK, DD-DD, KR-DK (7) \\ 
  2.0 & 2.0 & 10.5 & DD-KK, NK-NK, NN-NN, KR-KR, KK-KK, DD-DD, KR-ND, DN-NR (8) \\ 
  1.7 & 1.2 & 11.0 & DD-DD, KK-KK, DK-KD, DD-KR, KR-DD, NN-KD, KD-DK, NN-DR (8) \\
  1.2 & 1.2 & 12.0 & DK-KD, DD-DD, DR-KD, DR-DD, KK-KK, KR-KR, KR-DK (7) \\
  1.4 & 1.2 & 10.0 & ND-KR, NN-NN, KK-KK, DD-DD, DR-DR, KR-KR, NN-KD, DK-DR \\
   &  &  & NK-KN, NK-KR, DR-KN, KR-DK (12) \\
  1.4 & 1.4 & 11.5 & DD-KR, DD-DD, KK-KK, NN-NN, KR-DD, DR-NK, KR-ND (7) \\ \bottomrule
 \hline
\end{tabular}
\caption{A summary of the physical systems and corresponding model permutations studied.  We hold the observational uncertainty constant at 0.1mag for all light curves.  We denote the model permutations with the first letter for the dynamical ejecta and disk ejecta models used to generate the light curve, followed by a hyphen and the models provided to \texttt{PyMultiNest} as reference; thus, any permutations such as `AA-AA' or `AB-AB' correspond to direct sampling, whereas systems labeled e.g. `AA-BB' or `AB-CD'  correspond to cross sampling.  The total number of model permutations for each BNS is indicated in parentheses.}
\label{table:list_of_systems}
\end{table*}
\end{center}

\subsection{Parameter Estimation}

To sample our light curves, we use \texttt{PyMultiNest}, a Python adaptation of the \texttt{MultiNest} nested sampling algorithm developed by Buchner \cite{Buchner_2014}.  Nested sampling is a Bayesian inference method, distinct from Markov-Chain Monte Carlo (MCMC), that excels at comparing multiple models by `scoring' their relative ability to describe a dataset.  This is done via the Bayes factor, easily computed from the ratio of evidences for any two runs.  Nested sampling works by iteratively scoring the likelihoods of some fixed number of test points inside a volume of parameter space, and terminates once the evidence reaches some threshold value.  Thus, one can employ nested sampling to simultaneously obtain posterior distributions for a set of parameters as well as an overall measure of model performance (the evidence).  Starting with Bayes' Theorem, which describes the probability of some data $D$ being described by a given model $M$: 
\begin{equation}
Pr(M | D) = \frac{Pr(D | M)Pr(M)}{Pr(D)}
\end{equation}
where $Pr(D|M)$ is the conditional probability or likelihood that $M$ is true given $D$, $Pr(M)$ is the prior, and $Pr(D)$ is the evidence.  We assume uniform priors in $M_1,M_2$ and $R$; for mass, we consider $\mathcal{U}(1.0, 2.2)M_\odot$ and for radius, we have $\mathcal{U}(8.0,15.0)M_\odot$.  The details of how \texttt{PyMultiNest} works are provided in \cite{Buchner_2014}; a broader overview of nested sampling can be found in \cite{Buchner_2023}.

The likelihood function provided to \texttt{PyMultiNest} first calculates the tidal binary parameters ($C_{1,2}, \Lambda_{1,2}, \tilde{\Lambda}$) from the test points $M_1, M_2,$ and $R$.  Then, based on the specified dynamical and disk ejecta models, it obtains the total ejecta mass $M_{ej}$ and mean velocity $v_{ej}$, which serve as inputs for HN's light curve model.  The likelihood is then calculated through comparison of the generated light curve with an injected lightcurve (see below). We study two configurations, which we refer to as direct sampling and cross-sampling. In direct sampling, the injection and parameter estimation use the same mass models. In cross sampling, they use distinct mass models, to estimate biases due to mass model choices.
%; we cross-sample the light curves in our study by sampling the light curve with a different set of reference ejecta models than we use to create the injection.  We cross sample the light curves in our system for a more robust perspective of model dependence on kilonova PE.  
After the light curve is generated, it is compared with the injected light curve for specified observation times and photometric filters. The likelihood $\mathcal{L}$ for a given $(M_1, M_2, R)$ is then simply 
%the summed difference between the injection and the light curve calculated from the test points:
\begin{equation}
\textrm{log}_{10}(\mathcal{L}) = -0.5 \sum_{\nu} \sum_{\textrm{t}} \frac{(m_{\nu,\textrm{inj}}(t) - m_{\nu,\textrm{PE}}(t))^2}{\sigma_\textrm{PE}^2}
\end{equation}  
Here, $m_\nu(t)$ is the apparent magnitude at time $t$ in a given filter band $\nu$; `inj' indicates the injected light curve and `PE' refers to the light curve generated from a set of sample points in \texttt{PyMultiNest}.  We direct the reader to Hotokezaka \& Nakar 2019 \cite{Hotokezaka_2020} for a detailed explanation of how $m_\nu(t)$ is computed from the input parameters.  For the scope of this work, we consider a constant photometric uncertainty of $\sigma_\textrm{EM}=0.1$ mag, simulate our data as observed with ZTF $g,r,$ and $i$ bands; and evaluate our light curve from $t=0.2-6.4$ days post-merger. We note that the choice of $\sigma_\textrm{EM}$ would mostly impact the width of our posterior distributions, but not potential biases in mass models.
 
 We sampled 6 physical systems a total of 49 times, with an approximately equal amount of direct and cross sampling for each.  The parameters for the 6 mock BNSs are indicated in Table \ref{table:list_of_systems}. The direct sampling method allows us to determine degeneracies between the intrinsic binary parameters in the absence of modeling bias; the cross-sampling method allows us to study modeling biases.
 
  %While we do not reproduce the entirety of our results here for the sake of brevity, they {\AH will be} publicly available through Zenodo, and we feature the key results of our study here.  
 
%%%%%%%%%%%%%%%%%%%%%%%%%%%%%%%%%%%%%%
%%%%%%%%%%%%%%%% RESULTS %%%%%%%%%%%%%%%%
%%%%%%%%%%%%%%%%%%%%%%%%%%%%%%%%%%%%%%
\section{Results}\label{sec:results}

Here we present and discuss our findings from sampling our suite of mock light curves.  After \texttt{PyMultiNest} terminates, it returns a file containing approximately 5000 equally-weighted posterior points $(M_1,M_2,R)$.  Correlations between any two parameters, as well as biases in the posterior distributions, are easily visualized with corner plots of the posterior distributions overlaid with the injected values.  We are interested in measuring (i) how consistently the sampler is able to recover each intrinsic parameter given some ejecta model as reference and (ii) model dependence in this behavior, i.e. being able to recover parameters with one model versus another, or ``cross-sampling."  By measuring the consistency (or lack thereof) between different models, we can obtain a rough proxy measurement of their uncertainty.  We also observe the morphology of posterior distributions to assess parameter degeneracies within the limitations of our very simple light curve model, and compare them to degeneracies in GW observations.

\begin{figure*}
\includegraphics[width=7cm]{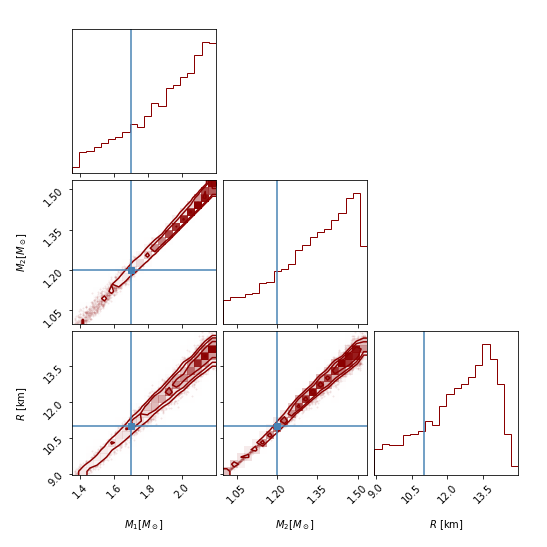} 
\includegraphics[width=7cm]{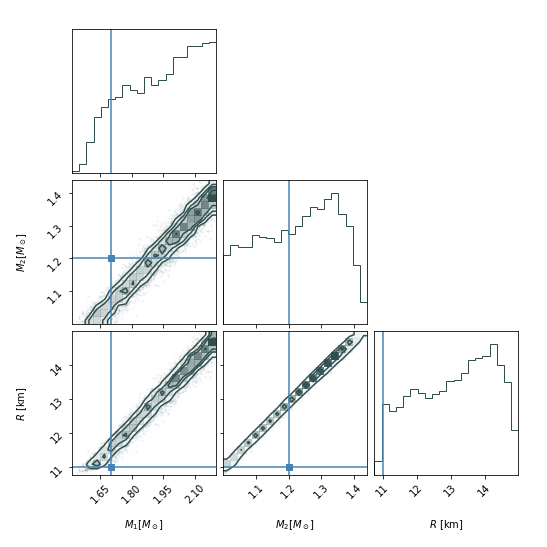}
\caption{Recovery of the intrinsic light curve parameters ($M_1$, $M_2$, and $R$) via \texttt{PyMultiNest} \cite{Buchner:2014nha} with true injection values $(M_1, M_2)=(1.7, 1.2)M_\odot, \textrm{R}=11$ km.  The light curve was produced using Dietrich \& Ujevic's dynamical ejecta model \cite{Dietrich_2017} and Dietrich \textit{et al.} 2020's \cite{Dietrich_2020} disk model, under the assumption that 30\% of the total disk mass is unbound as outflows.  We assume a constant error of 0.1 mag.  \textit{Left:} Sampling the light curve for $M_1$, $M_2$, and $R$ using the same dynamical and disk ejecta models as reference for our sampler (`direct sampling'). \textit{Right:} Sampling for the aforementioned parameters by providing \texttt{PyMultiNest} a different set of ejecta models as reference, namely Kruger \& Foucart 2020 \cite{Kruger_2020}'s dynamical ejecta model and the disk mass formula from Radice \textit{et al.} 2018 \cite{Radice_2018}.}
\label{fig:17m-12m-11km} % %%%%NOTE!!!%%%% this reference isn't working; i think it's because the figure has (*), but using it is the only way i've gotten the figure to span both columns
\end{figure*}

\begin{figure*}
\includegraphics[width=12cm]{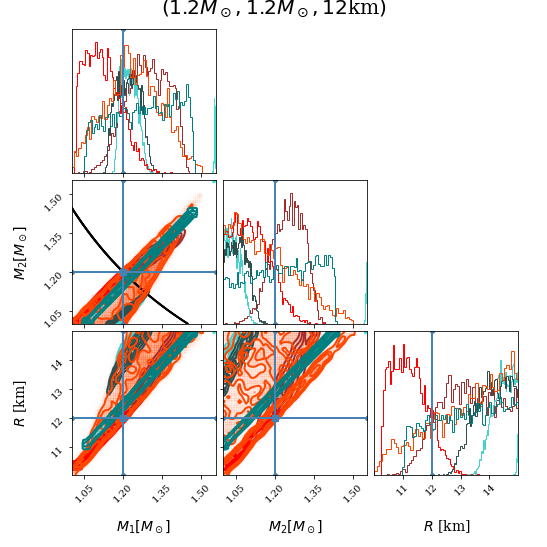}
\caption{The posterior distributions for $(M_1,M_2,R)=(1.2M_\odot, 1.2M_\odot, 12.0\textrm{km})$, with each model combination having its own contour.  For ease of visual reference, direct sampled systems are indicated in red tones whereas cross sampled posterior distributions are colored blue.  The line of constant chirp mass $M_c$ is indicated in black on the marginalized $M_1-M_2$ subplot.}
\label{fig:corner_1212}
\end{figure*}

\begin{figure*}
\includegraphics[width=12cm]{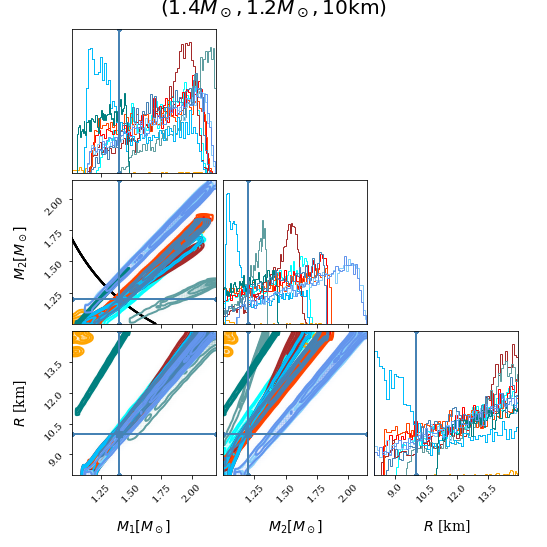}
\caption{The posterior distributions for $(M_1,M_2,R)=(1.4M_\odot, 1.2M_\odot, 10.0\textrm{km})$, with each model combination having its own contour.  Blue indicates cross sampled iterations; red corresponds to direct sampled iterations.  The line of constant $M_c$ is colored in black on the $M_1-M_2$ subplot.}
\label{fig:corner_1412}
\end{figure*}

\begin{figure*}
\includegraphics[width=12cm]{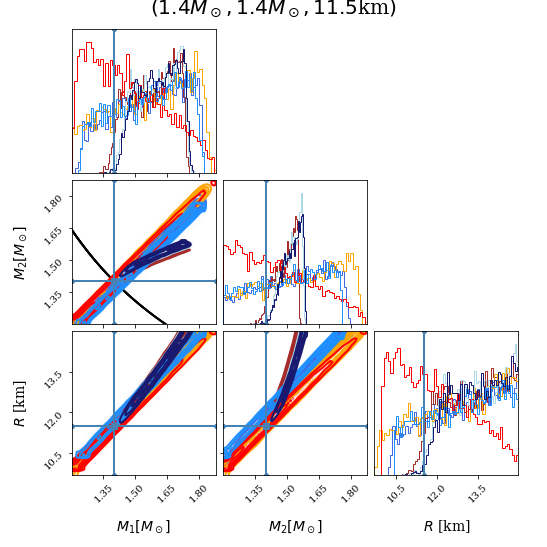}
\caption{The posterior distributions for $(M_1,M_2,R)=(1.4M_\odot, 1.4M_\odot, 11.5\textrm{km})$, with each model combination having its own contour.  Blue indicates cross sampled iterations; red corresponds to direct sampled iterations.  The line of constant $M_c$ is colored in black on the $M_1-M_2$ subplot.}
\label{fig:corner_1414}
\end{figure*}

\begin{figure*}
\includegraphics[width=12cm]{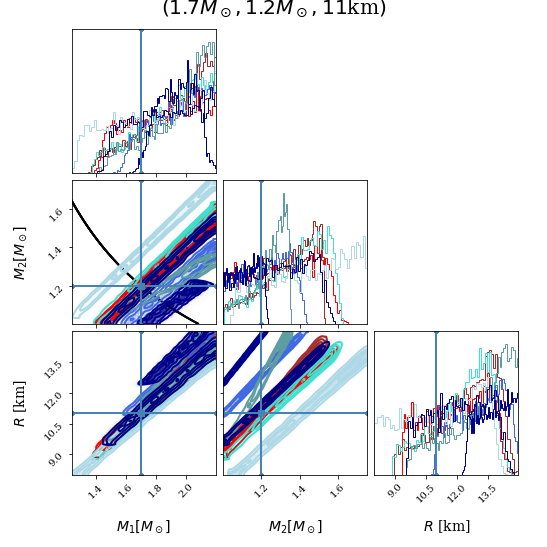}
\caption{The posterior distributions for $(M_1,M_2,R)=(1.7M_\odot, 1.2M_\odot, 11.0\textrm{km})$, with each model combination having its own contour.  Blue indicates cross sampled iterations; red corresponds to direct sampled iterations.  The line of constant $M_c$ is colored in black on the $M_1-M_2$ subplot.}
\label{fig:corner_1712}
\end{figure*}

\begin{figure*}
\includegraphics[width=12cm]{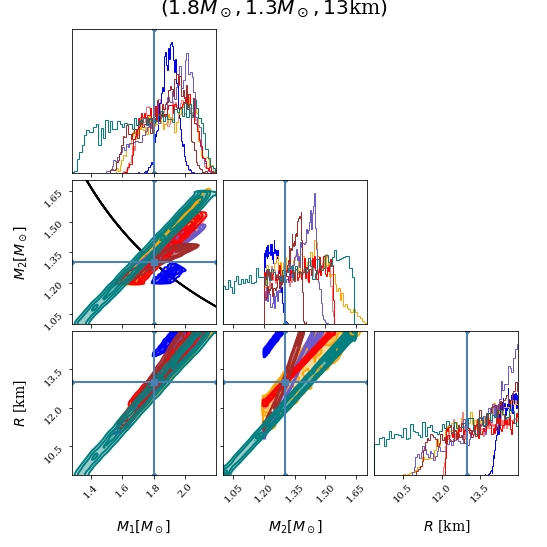}
\caption{The posterior distributions for $(M_1,M_2,R)=(1.8M_\odot, 1.3M_\odot, 13.0\textrm{km})$, with each physical system its own contour.  For ease of visual reference, direct sampled systems are indicated in red tones whereas cross sampled posterior distributions are colored blue.  Lines of constant chirp mass $M_c$ are indicated in black on the marginalized $M_1-M_2$ subplots for each BNS.}
\label{fig:corner_1813}
\end{figure*}

\begin{figure*}
\includegraphics[width=12cm]{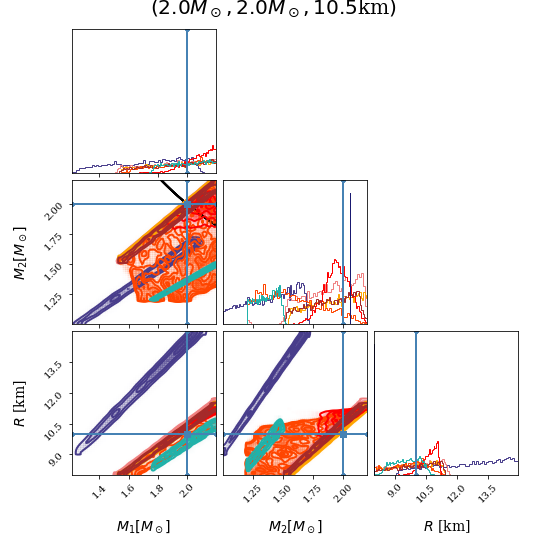}
\caption{The posterior distributions for $(M_1,M_2,R)=(2.0M_\odot, 2.0M_\odot, 10.5\textrm{km})$, with each physical system its own contour.  For ease of visual reference, direct sampled systems are indicated in red tones whereas cross sampled posterior distributions are colored blue.  Lines of constant chirp mass $M_c$ are indicated in black on the marginalized $M_1-M_2$ subplots for each BNS.}
\label{fig:corner_2020}
\end{figure*}
%Figures \ref{fig:corner_1212,fig:corner_1412,fig:corner_1414,fig:corner_1712,fig:corner_1813,fig:corner_2020},

In our corner plots, we include the injection values of each parameter in blue.  Above each one-dimensional histogram is the respective parameters' median and $1\sigma$ errors.  We regard a posterior distribution as having a \textit{bias} if the injection does not fall within the posterior.  
We also include lines of constant chirp mass $M_c$ (calculated from the injected $M_1$ and $M_2$) on the marginalized $M_1-M_2$ contours.  For candidate GW events, $M_c$ is typically the parameter measured with the most precision as it governs waveform evolution to first order. Comparing the $M_1-M_2$ contours to lines of constant $M_c$ enables us to identify the gain in information from a coincident EM detection.  A posterior which `cuts' through the $M_c$ line orthogonally would be maximally informative in identifying the system's component masses.  Conversely, if the posterior and $M_c$ were to intersect across the full mass range, we would be unable to exclude any $(M_1,M_2)$ pair along the curve and therefore gain no additional information.   

To aide our discussion, we include corner plots of the posterior distributions for one direct sampling and one cross sampling instance of $(1.7M_{\odot}, 1.2M_{\odot}, 11.0 \textrm{km})$ in Figure \ref{fig:17m-12m-11km}.  For both sets of results, the light curve is generated with DU's dynamical ejecta model and DEA's disk model (`DD'); the direct sampling instance employs the same models as reference for PE, whereas the cross sampled case employs KF's and REA's dynamical and disk ejecta models, respectively.  From Figure \ref{fig:17m-12m-11km}, we see that the direct sampled (left) system's posterior includes the injection, whereas the cross sampled (right) system is biased.  This indicates model disagreement between KR and DD in the specified region of parameter space $(1.7M_\odot, 1.2M_\odot, 11.0\textrm{km})$.  Specifically, looking at the distance between the injected values and the $90\%$ credibility region of the posterior probability in corner plots, we see biases of $O(0.1M)\odot$ in the masses and $O(1\,{\rm km})$ in the radius -- though due to the significant degeneracies between $(M_1,M_2,R)$, this bias is largely invisible in the marginalized one-dimensional posteriors. The differences between models are much clearer in the 2D posteriors than in the 1D posteriors, which is commonly the case in our results.   

We present an overlay of their posterior contours in Figures \ref{fig:corner_1212},\ref{fig:corner_1412},\ref{fig:corner_1414},\ref{fig:corner_1712}\ref{fig:corner_1813}, and \ref{fig:corner_2020}.  To have produced a result consistent with the injection, any given morphology must pass through the crosshairs for each injection parameter.  Thus, \ref{fig:corner_1212},\ref{fig:corner_1412},\ref{fig:corner_1414},\ref{fig:corner_1712}\ref{fig:corner_1813}, and \ref{fig:corner_2020} easily visualize the biases present while simultaneously highlighting how different models compare in their recovery.

The marginalized $M_1-M_2$ distributions in Figures \ref{fig:corner_1212},\ref{fig:corner_1412},\ref{fig:corner_1414},\ref{fig:corner_1712},\ref{fig:corner_1813}, and \ref{fig:corner_2020} demonstrate varying levels of agreement that appears to correlate with total mass, with BNSs at the margins of our mass prior faring somewhat worse.  One cross-sampled system for $(1.2,1.2)M_\odot$ and one for $(2.0, 2.0)M_\odot$ even fully failed to converge, i.e. the log-likelihood and global log-evidence remained quite large for the duration of the sampling.  Such results are to be given very low credence, but we keep them in our results here to demonstrate where the models begin to diverge.  In the corner plots, the width along the injected values in either direction can be thought of as the uncertainty solely due to model choice.  %The unconverged runs in our data manifest as pointlike estimates biased from the injection in $M_1-M_2$ --- see Figures \ref{fig:overlay_corners} and \ref{fig:m1-m2-with-mc}.  
%{\AH 1.2-1.2 wasn't as bad as 2.0-2.0 non-convergence, but need to check}  %(Although we identify their non-convergence here in comparison with our other, better-behaved results, we note that this feature is also identifiable in isolation.)    %($\mathcal{O}(10^6)$ vs. $\mathcal{O}(\sim1-10)$ for converged results) 
Conversely, $(1.4, 1.4)M_\odot, (1.4, 1.2)M_\odot,$ and $(1.8, 1.3)M_\odot$ (Figures \ref{fig:corner_1414},\ref{fig:corner_1412},\ref{fig:corner_1813}) each have posteriors which (with one exception) are consistent with the injected values.  We observe a bias of only $\sim(0.05-0.1)M_\odot$ for the $M_1-M_2$ recovery of $(1.4,1.4)M_\odot$ \ref{fig:corner_1414}; for $(1.4, 1.2)M_\odot$ \ref{fig:corner_1412}, we find similar recoverability, but are limited by two outliers which increase the bias in mass recovery to approximately $0.2M_\odot$.  For $(1.8, 1.3)M_\odot$ \ref{fig:corner_1813}, the masses are recovered to within $\sim 0.1M_\odot$  For $(1.7, 1.2)M_\odot$ \ref{fig:corner_1712}, the injected $M_\textrm{ej}$ itself varies so significantly between models that the corresponding posterior distributions do not overlap in $M_1-M_2$ or $M_2-R$.  This manifests a bias of approximately $0.2M_\odot$ for the recovery of either injected mass due to model uncertainty.  \textbf{The variance shown between model permutations can be thought of as the amount of bias due to choice of model in the pipeline.}  
%{\FF I replaced uncertainty with bias in the above paragraph, because it seems that you are talking about distance between cross-sampled posteriors, and not uncertainty within each posterior distribution; i.e. the marginalized posteriors each have much larger uncertainty than what is discussed above. Please check though; the current figures are too hard to read for me to figure this out with any certainty}

We find that our PE results are unilaterally perpendicular to $M_c$ (with the exception of an unconverged result for $(2.0, 2.0)$ \ref{fig:corner_2020}), but do not unilaterally intersect with one another.  In the case of a coincident GW detection, constraints on the chirp mass would in turn significantly constrain EM measurements to only those intersecting with $M_c$.  At this point, the choice of ejecta model will become a non-negligible source of uncertainty, because it influences where along the line of $M_c$ the EM posterior will intersect.  
A larger variance is more prone to occurring where the predicted ejecta mass is itself highly variable (model-dependent).  
A more common feature we observe is the posteriors intersecting with $M_c$ \textit{away from} the injected parameters, which is generally another indicator of model bias, but can also suggest poor sampler convergence (if the global log-evidence is much larger in comparison to other model combinations). We verified that except for the two runs previously mentioned, the convergence of the solver is not the source of the observed biases. We additionally discuss in the Appendix other diagnostics confirming that the biases are due to the choice of mass models.

%For example, the distributions for $(1.7, 1.2)M_\odot$ each trace out a different total mass, while having a more or less constant morphology.  This can be contrasted to the distributions for $(1.4,1.4)M_\odot$, which all pass through the injected $(M_1,M_2)$ and are in strong agreement compared to the other BNSs.  Here, the width of any given posterior result is the main source of uncertainty, as opposed to the distributions being in disagreement, or spread across a large swatch of parameter space (such as for $(1.7, 1.2)M_\odot$ or $(2.0, 2.0)M_\odot$). 
%(A merger with prompt collapse is as unlikely as a $q\sim2$ merger to produce significant early dynamical ejecta, but due to very different physical mechanisms, for example.) \\
%{\FF Is it true that $q\sim2$ has little dynamical ejecta? I would have thought tidal ejecta could be significant then}

As a way to obtain a point-like estimate of \texttt{PyMultiNest}'s ``best-guess,'' we sort each posterior file by log-likelihood and pull the point with the log-likelihood closest to zero (likelihood closest to one).  Each run ends up identifying an area of parameter space with strong evidence and sampling tightly around that region for the last $\sim20\%$ of posterior points; the smallest log-likelihood point is, in each case, nearly identical to the average of the final 100 posterior points, which is how we justify its use. % {\FF I don't think that we can necessarily claim that considering the wide spread of the posteriors; it would still be best to use the highest ranked 100 points, as this is not hard to find and does not require hand-wavy justifications}

\begin{figure*}
\includegraphics[width=5cm]{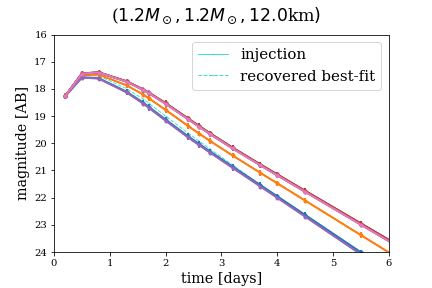} 
\includegraphics[width=5cm]{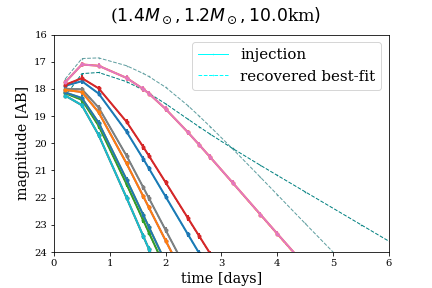}
\includegraphics[width=5cm]{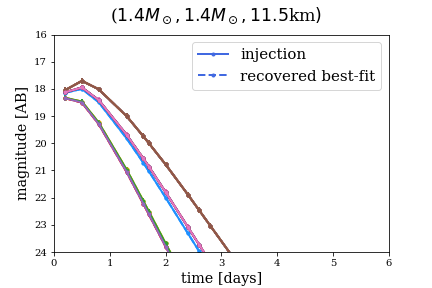} 
\includegraphics[width=5cm]{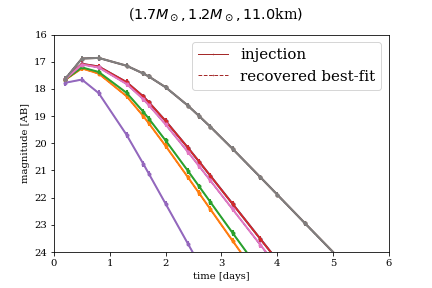}
\includegraphics[width=5cm]{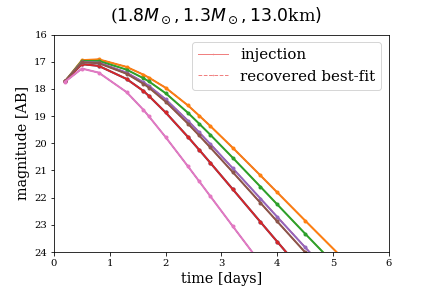} 
\includegraphics[width=5cm]{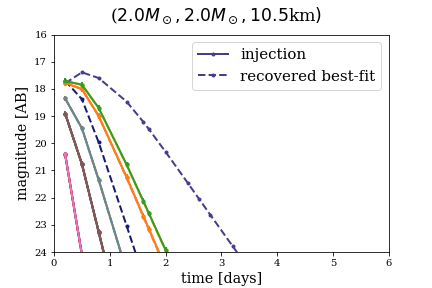}
\caption{Injected and recovered best-fit light curves for each model combination.  The six mock BNSs are each shown on separate panels.  The best-fit light curve is calculated from the lowest log-likelihood point in each respective run's posterior file.  Color-coded by model combination with red indicating direct sampling and blue for cross-sampling, each injection-recovery pair is shown in the same color, but with different linestyles (the injections are depicted with dashed lines).  Therefore, a good agreement with the injection is indicated by overlap between the same colored dashed and solid lines.  This occurs in all cases where $\texttt{PyMultiNest}$ is able to converge; see KR-ND $(2.0, 2.0)M_\odot$ (purple) for a counterexample, which terminated with a global log-evidence five orders of magnitude larger than the others in our sample.}
\label{fig:all_lcs}
\end{figure*}

As we have shown, different ejecta formulae can be used to obtain the same resultant mass (even if they map to different binary parameters), but can also vary significantly in the corresponding intrinsic binary parameters.  We wish to visualize how the different predicted $M_\textrm{ej}$ and $v_\textrm{ej}$'s impact the difference between resultant light curves, which we do in Figure \ref{fig:all_lcs}.  Again, we have color-coded the data according to choice of model for easier comparison.  

In Figure \ref{fig:all_lcs}, we demonstrate the full suite of model combinations for each physical system in one filter band (ZTF-$g$) to simultaneously show the baseline variance between injected light curves as well as the fidelity of our recovered light curves.  We find that the injection and best-fit recovered light curves agree to within our quoted errors of 0.1 mag as long as \texttt{PyMultiNest} is able to converge.  For example, we have a cross-sampled system for $(2.0, 2.0)M_\odot$ (green) which terminated with large likelihood, indicating poor quality-of-fit; this is made visually evident in Figure \ref{fig:all_lcs} (top left panel) by the large discrepancy between the injected and best-fit light curve.  
In general, the recovered light curve agrees strongly with the injection, but significant variations are observed depending on the model used for that injection.  Those variations differ between physical systems. Notably, $(1.2, 1.2)M_\odot$ and $(1.8, 1.3)M_\odot$ decay the slowest and feature the smallest amount of variance between different model predictions.  

\section{Conclusions}\label{sec:conclusion}
In this manuscript we have investigated how reliably various semi-analytical ejecta formulae perform in a PE context to relate kilonova observations to intrinsic BNS properties.  With simplifying assumptions on the NS EOS, observational error, and ejecta geometry, we isolate the impact of ejecta model choice in a parameter estimation pipeline in GEMMA. We remind the reader that these ejecta models are themselves imperfect, in that they are each based off of slightly different sets of simulations, which themselves vary in their microphysics and physical assumptions.  As the impact of ejecta formula is isolated, the results of this work should be generalizable to other BNS pipelines.  We find that the choice of ejecta model used to infer properties of a candidate event is non-trivial, as the models disagree in much of the parameter space, particularly for the more “extreme” masses which lie at the edge of our priors ($(1.2, 1.2)M_\odot$ and $(2.0, 2.0)M_\odot$ both have models which either fail to converge or converge poorly).  The choice of ejecta model used in one’s PE pipeline will affect inferences about the amount of mass present.  For $(1.4, 1.4)M_\odot$, the PE results agree with one another quite well and are the most self-consistent within our data set.  For other systems, biases of a few tenths of a solar mass in the masses and $O(1\,{\rm km})$ in radius are commonly observed.   %; however, the ejecta formulae already agree with one another in this part of the parameter space.  
%{\FF Is the 1.4-1.2 actually good? It looks bad on Fig 3; and has a few models with serious biases on Fig 2} 
We thus recommend that ejected mass be estimated with multiple models when performing parameter estimation for an estimate of how much error is introduced for the desired set of parameters.  %%, even if only one model is included in an analysis pipeline,
We also verified that the observed biases are due, within our pipeline, to the use of different mass models. Even when using cross-sampling we properly recover the injected mass and velocity of the ejecta, and generally find best fit light-curves matching our injections, as expected.

Finally, we note that the posteriors obtained in this work are largely orthogonal to constant chirp mass lines, which indicates the potential of  EM observations to provide a significant amount of additional information for events with joint GW-EM measurements, if our ejecta models can be improved enough to trust EM inference.

\section{Acknowledgements}
\begin{itemize}
    \item A.H. and F.F. gratefully acknowledge financial support from the DOE through grant DE-SC0020435, and from NASA through grant 80NSSC18K0565.
    \item S.M. acknowledges financial support from the European Research Council (ERC) starting grant 'GIGA' (grant agreement number: 101116134).
\end{itemize}

\bibliography{bibliography}

\begin{thebibliography}{10}

\bibitem{Fong_2015}
W.~{Fong}, E.~{Berger}, R.~{Margutti}, and B.~A. {Zauderer}, ``{A Decade of Short-duration Gamma-Ray Burst Broadband Afterglows: Energetics, Circumburst Densities, and Jet Opening Angles},'' {\em \apj}, vol.~815, p.~102, Dec. 2015.

\bibitem{Troja_2018}
E.~Troja, G.~Ryan, L.~Piro, H.~van Eerten, S.~B. Cenko, Y.~Yoon, S.-K. Lee, M.~Im, T.~Sakamoto, P.~Gatkine, A.~Kutyrev, and S.~Veilleux, ``A luminous blue kilonova and an off-axis jet from a compact binary merger at z=0.1341,'' {\em Nature Communications}, vol.~9, Oct. 2018.

\bibitem{Rastinejad_2022}
J.~C. Rastinejad, B.~P. Gompertz, A.~J. Levan, W.-f. Fong, M.~Nicholl, G.~P. Lamb, D.~B. Malesani, A.~E. Nugent, S.~R. Oates, N.~R. Tanvir, A.~de~Ugarte~Postigo, C.~D. Kilpatrick, C.~J. Moore, B.~D. Metzger, M.~E. Ravasio, A.~Rossi, G.~Schroeder, J.~Jencson, D.~J. Sand, N.~Smith, J.~F.~A. Fernández, E.~Berger, P.~K. Blanchard, R.~Chornock, B.~E. Cobb, M.~De~Pasquale, J.~P.~U. Fynbo, L.~Izzo, D.~A. Kann, T.~Laskar, E.~Marini, K.~Paterson, A.~R. Escorial, H.~M. Sears, and C.~C. Thöne, ``A kilonova following a long-duration gamma-ray burst at 350 mpc,'' {\em Nature}, vol.~612, p.~223–227, Dec. 2022.

\bibitem{Levan_2023}
A.~J. Levan, B.~P. Gompertz, O.~S. Salafia, M.~Bulla, E.~Burns, K.~Hotokezaka, L.~Izzo, G.~P. Lamb, D.~B. Malesani, S.~R. Oates, M.~E. Ravasio, A.~Rouco~Escorial, B.~Schneider, N.~Sarin, S.~Schulze, N.~R. Tanvir, K.~Ackley, G.~Anderson, G.~B. Brammer, L.~Christensen, V.~S. Dhillon, P.~A. Evans, M.~Fausnaugh, W.-f. Fong, A.~S. Fruchter, C.~Fryer, J.~P.~U. Fynbo, N.~Gaspari, K.~E. Heintz, J.~Hjorth, J.~A. Kennea, M.~R. Kennedy, T.~Laskar, G.~Leloudas, I.~Mandel, A.~Martin-Carrillo, B.~D. Metzger, M.~Nicholl, A.~Nugent, J.~T. Palmerio, G.~Pugliese, J.~Rastinejad, L.~Rhodes, A.~Rossi, A.~Saccardi, S.~J. Smartt, H.~F. Stevance, A.~Tohuvavohu, A.~van~der Horst, S.~D. Vergani, D.~Watson, T.~Barclay, K.~Bhirombhakdi, E.~Breedt, A.~A. Breeveld, A.~J. Brown, S.~Campana, A.~A. Chrimes, P.~D’Avanzo, V.~D’Elia, M.~De~Pasquale, M.~J. Dyer, D.~K. Galloway, J.~A. Garbutt, M.~J. Green, D.~H. Hartmann, P.~Jakobsson, P.~Kerry, C.~Kouveliotou, D.~Langeroodi, E.~Le~Floc’h, J.~K. Leung, S.~P. Littlefair, J.~Munday,
  P.~O’Brien, S.~G. Parsons, I.~Pelisoli, D.~I. Sahman, R.~Salvaterra, B.~Sbarufatti, D.~Steeghs, G.~Tagliaferri, C.~C. Thöne, A.~de~Ugarte~Postigo, and D.~A. Kann, ``Heavy-element production in a compact object merger observed by jwst,'' {\em Nature}, vol.~626, p.~737–741, Oct. 2023.

\bibitem{Hamidani_2024}
H.~Hamidani, M.~Tanaka, S.~S. Kimura, G.~P. Lamb, and K.~Kawaguchi, ``Grb 211211a: The case for an engine powered over r-process powered blue kilonova,'' 2024.

\bibitem{Stratta_2025}
G.~Stratta, A.~M. Nicuesa~Guelbenzu, S.~Klose, A.~Rossi, P.~Singh, E.~Palazzi, C.~Guidorzi, A.~Camisasca, S.~Bernuzzi, A.~Rau, M.~Bulla, F.~Ragosta, E.~Maiorano, and D.~Paris, ``The puzzling long grb 191019a: Evidence for kilonova light,'' {\em The Astrophysical Journal}, vol.~979, p.~159, Jan. 2025.

\bibitem{Gottlieb_2023}
O.~Gottlieb, B.~Metzger, E.~Quataert, D.~Issa, T.~Martineau, F.~Foucart, M.~Duez, L.~Kidder, H.~Pfeiffer, and M.~Scheel, ``A unified picture of short and long gamma-ray bursts from compact binary mergers,'' 2023.

\bibitem{Radice_2018}
D.~Radice, A.~Perego, K.~Hotokezaka, S.~A. Fromm, S.~Bernuzzi, and L.~F. Roberts, ``Binary neutron star mergers: Mass ejection, electromagnetic counterparts, and nucleosynthesis,'' {\em The Astrophysical Journal}, vol.~869, p.~130, Dec. 2018.

\bibitem{Gillanders_2023}
J.~H. Gillanders, E.~Troja, C.~L. Fryer, M.~Ristic, B.~O'Connor, C.~J. Fontes, Y.-H. Yang, N.~Domoto, S.~Rahmouni, M.~Tanaka, O.~D. Fox, and S.~Dichiara, ``Heavy element nucleosynthesis associated with a gamma-ray burst,'' 2023.

\bibitem{Kedia_2023}
A.~Kedia, M.~Ristic, R.~O’Shaughnessy, A.~B. Yelikar, R.~T. Wollaeger, O.~Korobkin, E.~A. Chase, C.~L. Fryer, and C.~J. Fontes, ``Surrogate light curve models for kilonovae with comprehensive wind ejecta outflows and parameter estimation for at2017gfo,'' {\em Physical Review Research}, vol.~5, Mar. 2023.

\bibitem{Ristic_2025}
M.~Ristić, R.~O'Shaughnessy, K.~Wagner, C.~J. Fontes, C.~L. Fryer, O.~Korobkin, M.~R. Mumpower, and R.~T. Wollaeger, ``Joint electromagnetic and gravitational wave inference of binary neutron star merger gw170817 using forward-modeling ejecta predictions,'' 2025.

\bibitem{Kruger_2020}
C.~J. Krüger and F.~Foucart, ``Estimates for disk and ejecta masses produced in compact binary mergers,'' {\em Physical Review D}, vol.~101, May 2020.

\bibitem{Dietrich_2017}
T.~Dietrich and M.~Ujevic, ``Modeling dynamical ejecta from binary neutron star mergers and implications for electromagnetic counterparts,'' {\em Classical and Quantum Gravity}, vol.~34, p.~105014, Apr. 2017.

\bibitem{Nedora_2021}
V.~Nedora, F.~Schianchi, S.~Bernuzzi, D.~Radice, B.~Daszuta, A.~Endrizzi, A.~Perego, A.~Prakash, and F.~Zappa, ``Mapping dynamical ejecta and disk masses from numerical relativity simulations of neutron star mergers,'' {\em Classical and Quantum Gravity}, vol.~39, p.~015008, Dec. 2021.

\bibitem{Henkel_2023}
A.~Henkel, F.~Foucart, G.~Raaijmakers, and S.~Nissanke, ``Study of the agreement between binary neutron star ejecta models derived from numerical relativity simulations,'' {\em Physical Review D}, vol.~107, Mar. 2023.

\bibitem{Korobkin_2021}
O.~Korobkin, R.~T. Wollaeger, C.~L. Fryer, A.~L. Hungerford, S.~Rosswog, C.~J. Fontes, M.~R. Mumpower, E.~A. Chase, W.~P. Even, J.~Miller, G.~W. Misch, and J.~Lippuner, ``Axisymmetric radiative transfer models of kilonovae,'' {\em The Astrophysical Journal}, vol.~910, p.~116, Apr. 2021.

\bibitem{Darbha_2020}
S.~Darbha and D.~Kasen, ``Inclination dependence of kilonova light curves from globally aspherical geometries,'' {\em The Astrophysical Journal}, vol.~897, p.~150, July 2020.

\bibitem{Zhu_2022}
J.-P. Zhu, Y.-P. Yang, B.~Zhang, H.~Gao, and Y.-W. Yu, ``Kilonova and optical afterglow from binary neutron star mergers. i. luminosity function and color evolution,'' 2022.

\bibitem{Ricigliano:2024lwf}
G.~Ricigliano, M.~Jacobi, and A.~Arcones, ``{Impact of nuclear matter properties on the nucleosynthesis and the kilonova from binary neutron star merger ejecta},'' {\em Mon. Not. Roy. Astron. Soc.}, vol.~533, no.~2, pp.~2096--2112, 2024.

\bibitem{Wollaeger_2018}
R.~T. Wollaeger, O.~Korobkin, C.~J. Fontes, S.~K. Rosswog, W.~P. Even, C.~L. Fryer, J.~Sollerman, A.~L. Hungerford, D.~R. van Rossum, and A.~B. Wollaber, ``Impact of ejecta morphology and composition on the electromagnetic signatures of neutron star mergers,'' {\em Monthly Notices of the Royal Astronomical Society}, vol.~478, p.~3298–3334, Apr. 2018.

\bibitem{Suleiman_2021}
L.~Suleiman, M.~Fortin, J.~L. Zdunik, and P.~Haensel, ``Influence of the crust on the neutron star macrophysical quantities and universal relations,'' {\em Physical Review C}, vol.~104, July 2021.

\bibitem{Christie:2019lim}
I.~M. Christie, A.~Lalakos, A.~Tchekhovskoy, R.~Fern\'andez, F.~Foucart, E.~Quataert, and D.~Kasen, ``{The Role of Magnetic Field Geometry in the Evolution of Neutron Star Merger Accretion Discs},'' {\em Mon. Not. Roy. Astron. Soc.}, vol.~490, no.~4, pp.~4811--4825, 2019.

\bibitem{Metzger:2014ila}
B.~D. Metzger and R.~Fern\'andez, ``{Red or blue? A potential kilonova imprint of the delay until black hole formation following a neutron star merger},'' {\em Mon. Not. Roy. Astron. Soc.}, vol.~441, pp.~3444--3453, 2014.

\bibitem{Dietrich_2020}
T.~Dietrich, M.~W. Coughlin, P.~T.~H. Pang, M.~Bulla, J.~Heinzel, L.~Issa, I.~Tews, and S.~Antier, ``Multimessenger constraints on the neutron-star equation of state and the hubble constant,'' {\em Science}, vol.~370, p.~1450–1453, Dec. 2020.

\bibitem{Hotokezaka_2020}
K.~Hotokezaka and E.~Nakar, ``Radioactive heating rate of r-process elements and macronova light curve,'' {\em The Astrophysical Journal}, vol.~891, p.~152, Mar. 2020.

\bibitem{Raaijmakers_2021}
G.~Raaijmakers, S.~Nissanke, F.~Foucart, M.~M. Kasliwal, M.~Bulla, R.~Fernández, A.~Henkel, T.~Hinderer, K.~Hotokezaka, K.~Lukošiūtė, T.~Venumadhav, S.~Antier, M.~W. Coughlin, T.~Dietrich, and T.~D.~P. Edwards, ``The challenges ahead for multimessenger analyses of gravitational waves and kilonova: A case study on gw190425,'' {\em The Astrophysical Journal}, vol.~922, p.~269, Dec. 2021.

\bibitem{bilby_paper}
G.~Ashton {\em et~al.}, ``{BILBY: A user-friendly Bayesian inference library for gravitational-wave astronomy},'' {\em Astrophys. J. Suppl.}, vol.~241, no.~2, p.~27, 2019.

\bibitem{Buchner_2014}
J.~Buchner, A.~Georgakakis, K.~Nandra, L.~Hsu, C.~Rangel, M.~Brightman, A.~Merloni, M.~Salvato, J.~Donley, and D.~Kocevski, ``X-ray spectral modelling of the agn obscuring region in the cdfs: Bayesian model selection and catalogue,'' {\em Astronomy \& Astrophysics}, vol.~564, p.~A125, Apr. 2014.

\bibitem{Buchner_2023}
J.~Buchner, ``Nested sampling methods,'' {\em Statistics Surveys}, vol.~17, Jan. 2023.

\bibitem{Buchner:2014nha}
J.~Buchner, A.~Georgakakis, K.~Nandra, L.~Hsu, C.~Rangel, M.~Brightman, A.~Merloni, M.~Salvato, J.~Donley, and D.~Kocevski, ``{X-ray spectral modelling of the AGN obscuring region in the CDFS: Bayesian model selection and catalogue},'' {\em Astron. Astrophys.}, vol.~564, p.~A125, 2014.

\end{thebibliography}
\bibliographystyle{ieeetr}

\appendix

\section{}

To assist in visualizing how model dependence manifests in each part of our parameter estimation, we plot (i) the injected $(M_1,M_2,R)$ against the best-fit $(M_1,M_2,R)$ for each sampler run; (ii) the injected and best-fit ejecta parameters $(M_{\textrm{ej}},v_{\textrm{ej}})$; and (iii) the corresponding light curves.  In the switch from $(M_1,M_2,R)$ to $(M_{\textrm{ej}},v{\textrm{ej}})$, we introduce a degeneracy by virtue of reducing the dimensionality, and vary the relations used to obtain $M_{\textrm{ej}}$.  Then, the $(M_{\textrm{ej}},v_{\textrm{ej}})$ is fed to our light curve model, which is a non-linear function evolved in time; thus, some degeneracy is introduced here because the relationship is not straightforward or one-to-one. 
\begin{figure*}
\includegraphics[width=7cm]{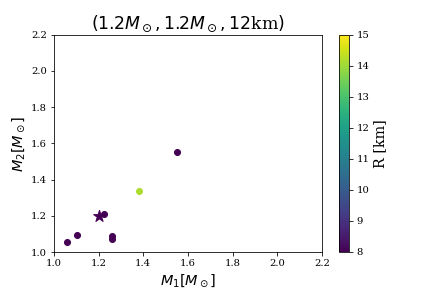}
\includegraphics[width=7cm]{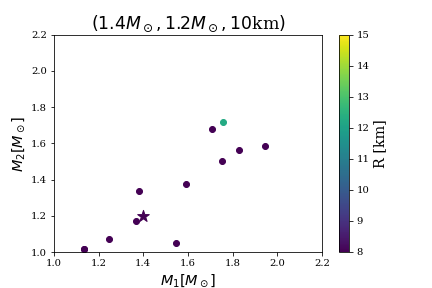}
\includegraphics[width=7cm]{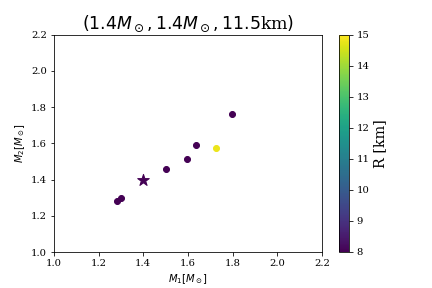}
\includegraphics[width=7cm]{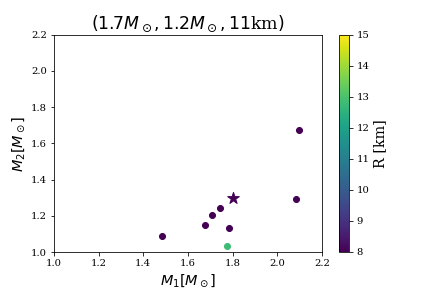}
\includegraphics[width=7cm]{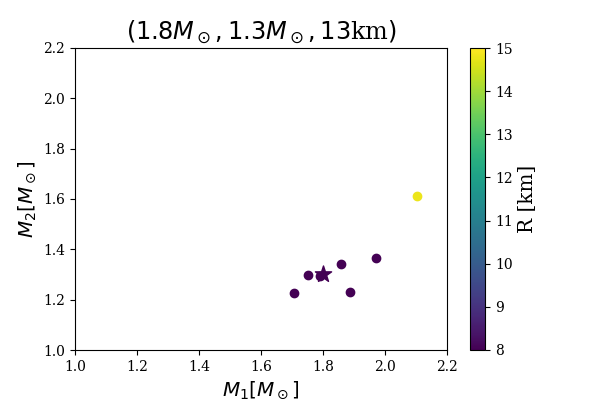}
\includegraphics[width=7cm]{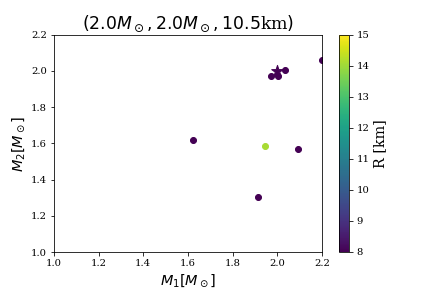}
\caption{The injected versus recovered best-fit $(M_1,M_2,R)$ for each BNS in this work, obtained from the highest-ranked (best-scored likelihood) posterior point from each posterior distribution.  The injection value is indicated above each subfigure as well as plotted as a star within the figures.} 
\label{fig:m1m2r}
\end{figure*}

We demonstrate the injected versus recovered best-fit $(M_1,M_2,R)$ in Figure \ref{fig:m1m2r}, which lies on the $M_1-M_2$ plane with color indicating radius.  From Figure \ref{fig:m1m2r}, we wish to emphasize the amount of variance in recovered best-fit masses and radii.  Notably, none of the best-fit $(M_1,M_2,R)$ are localized to the correct injection.  While the distributions for $(1.4, 1.2)M_\odot$ and $(1.4, 1.4)M_\odot$ preferentially overestimate the component masses and radius, this trend is not common to all BNSs analyzed here.  We contextualize this figure with the reminder that either neutron star mass or radius is not actually provided to the light curve model; rather, only the $M_{ej}$ and $v_{ej}$ are explicit functions of the kilonova signal. 

\begin{figure*}
\includegraphics[width=5cm]{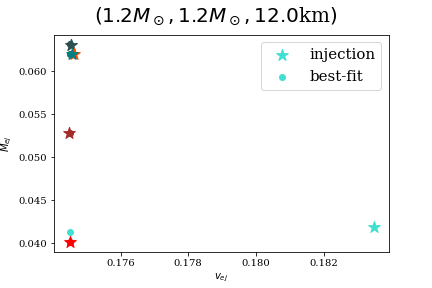}
\includegraphics[width=5cm]{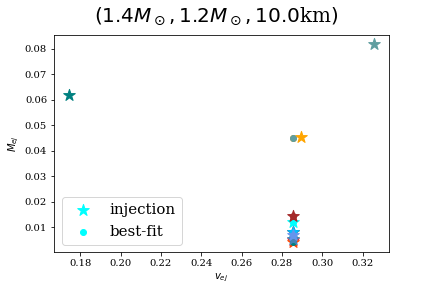}
\includegraphics[width=5cm]{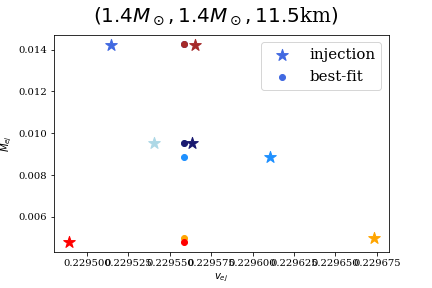}
\includegraphics[width=5cm]{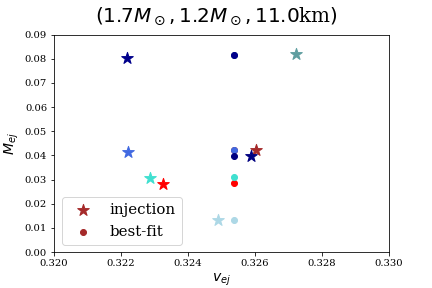}
\includegraphics[width=5cm]{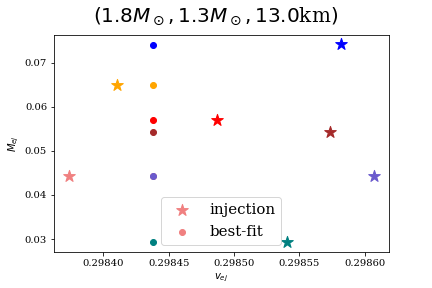}
\includegraphics[width=5cm]{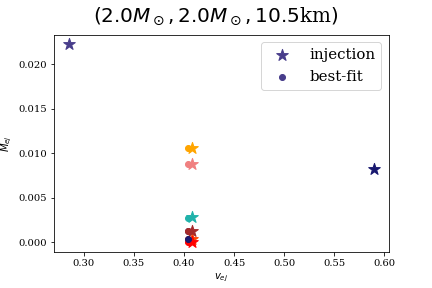}
\caption{For each BNS, the injected (star) versus recovered best-fit (dot) $(M_\textrm{ej}, v_\textrm{ej})$; each injection-recovery pair is shaded with the same color.}
\label{fig:mejvej}
\end{figure*}

 %Rather, we can deduce that the injected $(M_1,M_2,R)$ has little-to-no influence on the predicted $(M_1,M_2,R)$.  
In Figure \ref{fig:mejvej}, we show the injected versus best-fit $(M_\textrm{ej}, v_\textrm{ej})$ for each BNS, color-coded by model combination to better distinguish injection-recovery pairs.  Because we only employ one ejecta velocity model, the various $M_\textrm{ej}$ for direct sampling (star symbols) all have the same velocity.  Here, the ``width" along the $y-$ ($M_\textrm{ej}$-) axis between the topmost and lowest star can be regarded as the variance in $M_\textrm{ej}$ \textit{solely} due to choice of model, as these quantities are directly calculated from the injection parameters.  From this, we can see that there is a spread of $\sim0.01-0.1M_\odot$ between different ejecta formulae, depending on the system.  We find that the ejecta masses are most tightly constrained for $(2.0,2.0)M_\odot$, which has $0.01M_\odot$ of variance; here, all models agree that no significant amount of ejecta will be emitted.  However, even for this regime, the models vary from one another on the order of $0.01M_\odot$.
All cross-sampling results obtained from the same injected light curve result in nearly identical best-fit ejecta mass and velocity, which demonstrates that the sampler is able to very accurately identify the correct ejecta properties.  From these figures, we wish to highlight that although the masses and radii are not necessarily well-recovered, they still link closely to the observed amount of \textit{ejected} mass.  In each case, the limiting factor in ascertaining the ejected mass is the inherent variance due to choice of model, rather than sampler performance or the actual recoverability of the parameters of the ejecta.

%\end{appendices}

\end{document}